\documentclass[useAMS,usenatbib]{mn2e}
\usepackage{graphicx,multirow}
\def\mcg6{MCG--6-30-15}
\def\mrk7{Mrk\,766}
\def\1h0{1H\,0707-495}
\def\araa{ARA\&A}%
\def\apj{ApJ}%
\def\apjl{ApJ}%
\def\aap{A\&A}%
\def\mnras{MNRAS}%
\def\nat{Nature}%
\title[NXTL from \mcg6 and \mrk7]
{Negative X-ray reverberation time delays from \mcg6 and \mrk7\thanks{Based on observations obtained with \textit{XMM-Newton}, an ESA science mission with instruments and contributions directly funded by ESA Member States and NASA.}}
\author[D.~Emmanoulopoulos et al.]{D.~Emmanoulopoulos,$^{1}$\thanks{E-mail: D.Emmanoulopoulos@soton.ac.uk} I.~M.~M\textsuperscript{c}Hardy$^{1}$ and I.~E.~Papadakis,$^{2,3}$ \\
$^{1}$School of Physics and Astronomy, University of Southampton, SO17 1BJ Southampton, United Kingdom\\
$^{2}$Department of Physics and Institute of Theoretical \& Computational Physics, University of Crete, 71003 Heraklion, Greece\\
$^{3}$IESL, Foundation for Research and Technology, 71110 Heraklion, Greece}

\begin{document}
\date{Accepted 2011 June 29. Received 2011 June 24; in original form 2011 May 27}
\pagerange{\pageref{firstpage}--\pageref{lastpage}} \pubyear{2002}
\maketitle
\label{firstpage}

\begin{abstract}
We present an X-ray time lag analysis, as a function of Fourier frequency, for \mcg6 and \mrk7 using long term \textit{XMM-Newton} light curves in the 0.5--1.5 keV and the 2--4 keV energy bands, together with some physical modelling of the corresponding time lag spectra. Both the time lag spectra of \mcg6 and \mrk7 show negative values (i.e.\ soft band variations lag behind the corresponding hard band variations) at high frequencies, around $10^{-3}$ Hz, similar to those previously observed from \1h0. The remarkable morphological resemblance between the time lag spectra of \mcg6 and \mrk7 indicate that the physical processes responsible for the observed soft time delays is very similar in the two sources, favouring a reflection scenario from material situated very nearby to the central black hole. 
\end{abstract}

\begin{keywords}
galaxies: individual: X-rays: galaxies  -- galaxies: nuclei -- galaxies: Seyfert -- black hole physics
\end{keywords}

\section{Introduction}
\label{sect:intro}
The measurement of time lags as a function of Fourier frequency between soft and hard X-ray energy bands is widely used as a diagnostic of the emission mechanism and source geometry in active galactic nuclei (AGN) \citep[e.g.][]{papadakis01,mchardy04,arevalo06,arevalo08,sriram09} as well as X-ray binaries (XRBs) \citep[e.g.][]{miyamoto89,nowak96,nowak99}. Usually they have positive values i.e.\ hard band variations lag soft band variations, and their origin is still ambiguous. Although positive time lags are expected in standard comptonisation models \citep{nowak99}, they can also be caused by diffusive propagation of perturbations in the accretion flow \citep{kotov01}.  
\par
Tentative detections of negative X-ray time lags (NXTL) (i.e.\ soft band variations lag hard band variations) were firstly reported for the type $\rmn{I}$ Seyfert galaxies Mrk\,766 \citep{markowitz07} and Ark\,564 \citep{mchardy07}, the latter commented that they could be due to reflection by material very close to the black hole (BH). The only robust detection of NXTL, so far, came from \citet{fabian09} for the narrow-line type $\rmn{I}$ Seyfert galaxy \1h0. They detected NXTL of 30 s for frequencies greater than $6\times 10^{-4}$ Hz. At lower frequencies, positive time lags, as commonly seen in AGN, were also detected.\par
The origin of NXTL is, if anything, even less clear. \citet{fabian09} and \citet{zoghbi10,zoghbi11} favour the interpretation that they arise from reflection by matter lying very close to the BH. However, alternative explanations involving large scale distant reflectors with anisotropic geometry (e.g.\ accretion disc wind) can also reproduce them \citep{miller10_1h0707}. Thus, ideal NXTL candidates are the X-ray bright AGN which have long and continuous soft and hard X-ray light curves and which show evidence of reprocessing e.g.\ possessing broad X-ray iron emission lines or having strong and variable soft excesses in their X-ray spectra.\par
The type $\rmn{I}$ Seyfert galaxy \mcg6 was the first AGN reported to possess a broad X-ray iron line \citep{tanaka95}. Its reflection spectrum is considered archetypical for other AGN exhibiting similar broad iron line features \citep[e.g.][]{nandra07}. Based on the width of this line, using \textit{XMM-Newton} observations, \citet{wilms01} and \citet{fabian02} have put constraints for \mcg6 on the inner radius of the emission region of the order of 2 gravitational radii ($r_{\rm g}$). Another X-ray bright type $\rmn{I}$ Seyfert galaxy, observed repeatedly by \textit{XMM-Newton} is \mrk7. Although not one of the most convincing cases, it shows weak evidence of broad relativistic iron line \citep{miller06,miller07}.\par
In this letter we re-analyse the archival \textit{XMM-Newton} data of the above two sources and we report the existence of NXTL for both of them. Time lag analyses of the same data sets have been already performed by \citet{vaughan03b} for \mcg6 and \citet{markowitz07} for \mrk7. These analyses used large time bins (100 s and 60 s respectively) and were aimed to probe more the low frequencies part of the time lag spectrum (below $10^{-3}$ Hz). They measured positive time delays dropping reciprocally with the frequency. In addition both analyses used energy bands (0.2--0.7 vs. 0.7--2 and 0.2--0.5 vs. 0.5--1.1 respectively) not appropriate to enhance the existence of a possible reflection signature. In Section \ref{sect:data_red} we describe the data sets and the data reduction procedures, in Section \ref{sect:time_lag_estim} we estimate and model the time lag spectra, and finally in Section \ref{sect:discus} we discuss our results.

\section{DATA SETS AND DATA REDUCTION}
\label{sect:data_red}
\mcg6 has been observed six times and \mrk7 eight times by \textit{XMM-Newton} (all these data are publicly available). 
To achieve the most accurate determination of the time lag spectrum in both sources, we selected the observations which have an on-source exposure time larger than one full-day (i.e.\ more than 86.4 ks). The log for these observations is listed in Table~\ref{tab:data_sets}.\par
The EPIC raw-data for all the observations are reduced with the \textit{XMM-Newton} {\sc Scientific Analysis System} ({\sc SAS}) \citep{gabriel04} version 10.0.2. After reprocessing the pn and the two MOS data sets with the \textit{epchain} and the \textit{emchain}  {\sc SAS}-tools respectively, we perform a thorough check for pile-up using the task \textit{epatplot}. For the production of the light curves we select events that are detected up to quadruple pixel-pattern on the CCDs i.e.\ PATTERN$<$12. For each observation and for each instrument separately, the corrected background-subtracted light curves in the 0.5--1.5 keV and 2--4 keV energy ranges are produced using the {\sc SAS}-tool \textit{epiclccorr} in time-bins of 20 s. Then, for each observation and for each energy range we combine the background-subtracted light curves from pn, MOS\,1 and MOS\,2. Additionally, for each observation we also create the pn background light curve, in the 0.5--10 keV energy range in order to trace time-periods with increased background activity that are going to be disregarded from our analysis. Typically, after the abovementioned data reduction procedures the length of the final effective light curves are of the order of 0.5--20 per cent less with respect to the initial pointing on-source time (see Table~\ref{tab:data_sets} columns `On time' and `Eff.time').\par
Finally, for comparison, we also consider the `at-least one-day long' publicly available archival \textit{XMM-Newton} data for \1h0. These are the ones used by \citet{fabian09} and \citet{zoghbi10,zoghbi11} (see Table~\ref{tab:data_sets} for details).

\begin{table*}
\centering
\begin{minipage}{140mm}
\caption{Summary of the \textit{XMM-Newton} observations for all the sources.}
\label{tab:data_sets}
\begin{tabular}{@{}lcccccccc}
\hline
\multirow{2}{*}{Source} & \multirow{2}{*}{Obs. Id} & \multirow{2}{*}{Revolution} & \multirow{2}{*}{Start time} & \multirow{2}{*}{On time} & \multirow{2}{*}{Eff.time} & \multirow{2}{*}{Observing} & \multirow{2}{*}{Count rate} \\[0.3em] &  &  & (UTC) & (ks) & (ks)\footnote{The effective observing time, i.e.\ light curve duration, after the reduction of the raw data.} & mode\footnote{For the pn, MOS\,1, and MOS\,2 respectively. sw:small-window, lw:large window, and fw:fast-uncompressed.} & (s$^{-1}$)\footnote{The mean count rate and the standard deviation of the light curve in the 0.5--4 keV energy band.}\\
\hline
\multirow{3}{*}{\mcg6}  & 0029740101 & 0301 & 2001-08-01 16:07:44 & 89.4 & 81.3 & sw,sw,sw & 32.3$\pm$10.4 \\
& 0029740701 & 0302 & 2001-08-02 04:25:17 & 129.4 & 123.3 & sw,sw,sw & 35.9$\pm$7.9\\
& 0029740801 & 0303 & 2001-08-04 04:19:10  & 130.5 & 124.3  & sw,sw,sw & 32.9$\pm$13.5\\
\hline
\multirow{6}{*}{\mrk7} &  0109141301 & 0265 & 2001-05-20 08:33:34  & 129.9 & 113.6 & sw,fu,sw & 28.2$\pm$7.3 \\
& 0304030101 & 0999 & 2005-05-23 19:15:52 & 95.5 & 77.8 &  sw,lw,lw & 5.8$\pm$ 2.3\\
& 0304030301 & 1000 & 2005-05-25 18:16:53  & 98.9 & 98.3 &  sw,lw,lw & 11.8$\pm$4.6\\
& 0304030401 & 1001 & 2005-05-27 18:19:03  & 98.9 &  93.1 & sw,lw,lw & 15.2$\pm$4.2\\
& 0304030501 & 1002 & 2005-05-29 19:11:44 & 95.5 &  95.0 & sw,lw,lw & 18.6$\pm$4.5\\
& 0304030601 & 1003 & 2005-05-31 18:12:53 & 98.9 &  98.4 & sw,lw,lw & 15.1$\pm$5.4\\
\hline
\multirow{4}{*}{\1h0} & 0511580101 & 1491  & 2008-01-29 18:28:24 & 123.8 & 121.6 & lw,sw,sw & 3.0$\pm$1.3\\
& 0511580201 & 1492 & 2008-01-31 18:20:46  & 123.7 & 102.1 & lw,sw,sw & 4.6$\pm$1.6\\
& 0511580301 & 1493 & 2008-02-02 18:22:11  & 122.5 & 104.1 & lw,sw,sw & 4.1$\pm$1.7\\
& 0511580401 & 1494 & 2008-02-04 18:24:21  & 121.9 & 101.7 & lw,sw,sw & 3.2$\pm$1.4\\
\hline
\end{tabular}
\end{minipage}
\end{table*}

\section{Time lag estimation}
\label{sect:time_lag_estim}
For each source and for each observation we estimate, using the phases of the cross-spectrum \citep[e.g][]{priestley81}, the time lag spectrum, $\tau(f)$, at a given Fourier frequency, $f$, between the light curves in the soft, 0.5--1.5 keV, and hard, 2--4 keV, energy bands. In order to optimise the contribution from the reflection component or the scattering/absorbing material, we choose the 0.5--1.5 keV band where the `soft excess', above an extrapolation from the higher energy power law continuum, is greatest. Below 0.5 keV the response of the PN instruments is not well calibrated. For the hard band, where we aim to isolate the continuum emission, we choose 2--4 keV. This band is relatively free of contribution from the soft excess and also from the broad Fe K$\alpha$ line. Moreover, above 4 keV Poisson noise becomes significant contaminating further our data. During the estimation we average over a range of at least ten consecutive frequency bins and the corresponding error is computed following \citet{bendat86} and \citet{nowak99}. The same procedure is repeated for all the observations of the source, and we compute the average time lag value together with its error at the mean of the grouped frequency bins.\par
\begin{figure*}
\includegraphics[width=3in]{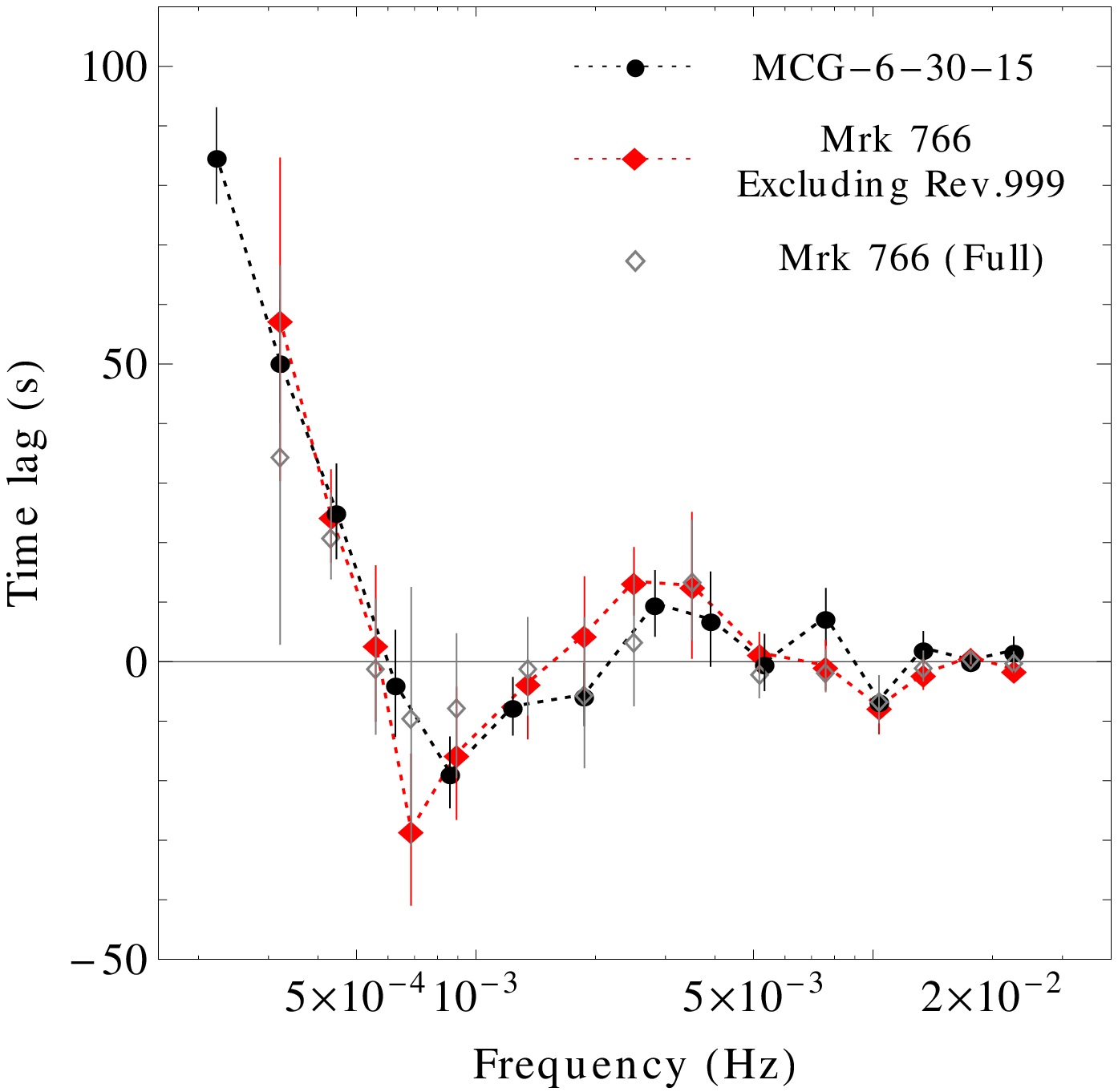}
\includegraphics[width=3in]{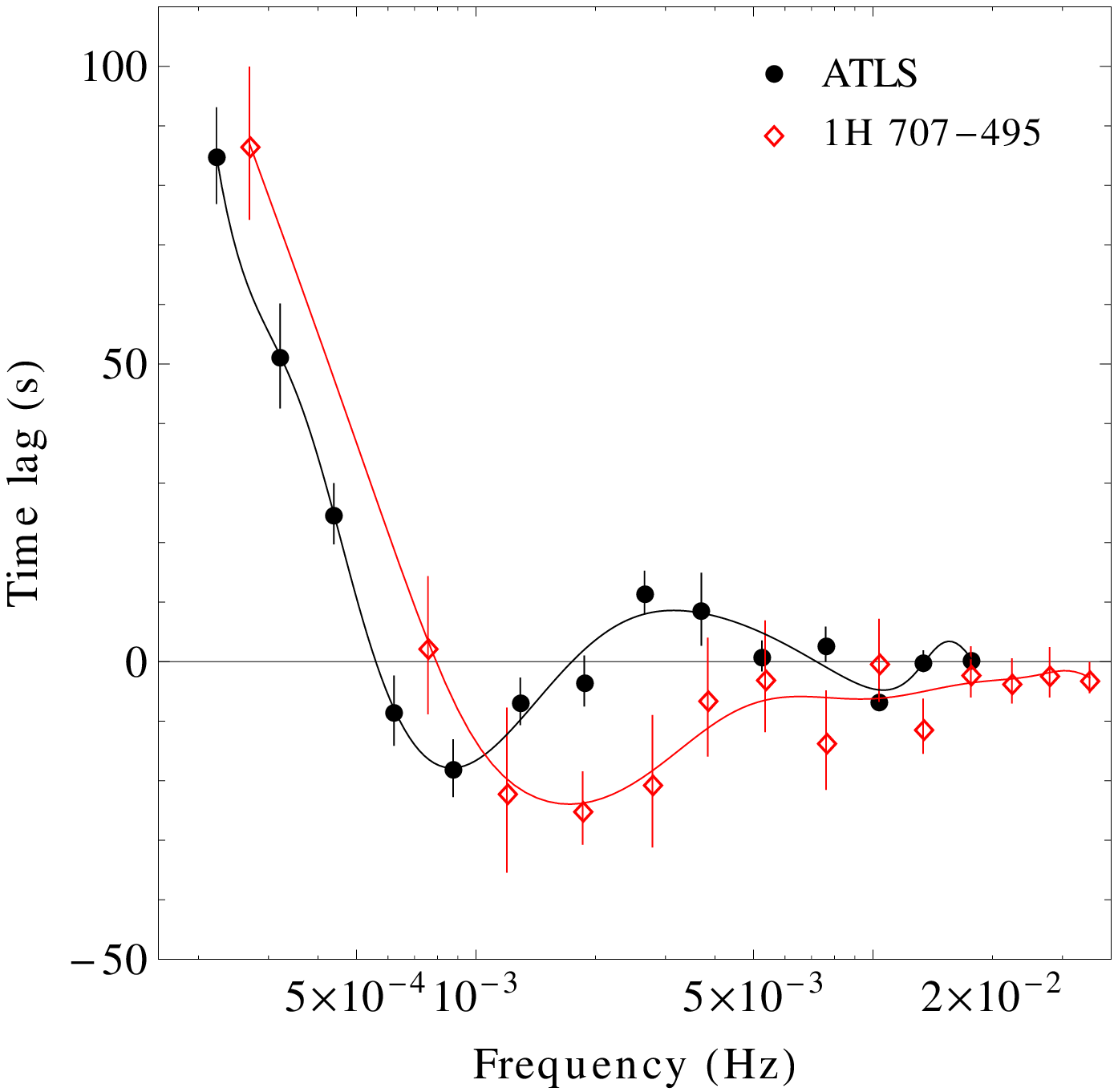}
\caption{Time lag spectra for \mcg6, \mrk7, and \1h0. Positive time lags indicate that variations in the 0.5--1.5 keV energy band lead the corresponding variations in the 2--4 keV energy band. (Left-hand panel). Time lag spectra for \mcg6 (black filled circles), \mrk7 [excluding revolution 999] (red filled diamonds), [all the observations] (grey open diamonds). The dotted lines among the points of the time lag spectra are linear interpolations intended to guide only the eye. (Right-hand panel) Average time lag spectrum (ATLS) of \mrk7 and \mcg6 (black filled circles) together with that of \1h0 (red open diamonds). The solid lines correspond to the best-fitting quadratic degree B-splines each one having ten control points.}
\label{fig:comb_time_lag} 
\end{figure*}

The time lag spectrum of \1h0 (Fig.~\ref{fig:comb_time_lag} right-hand panel red open-diamonds) has been already estimated by \citet{fabian09,zoghbi10,miller10_1h0707} and \citet{zoghbi11} and our result is in absolute accordance with them. Note that we use marginally different energy bands, with respect to the previous works cf. 0.3--1.0 keV vs.\ 1.0--4.0 keV.\par
In the left-hand panel of Fig.~\ref{fig:comb_time_lag} we show the estimated time lag spectra for \mcg6 (filled black-circles) and \mrk7 (open grey-diamonds) calculated using all of the data sets listed in Table~\ref{tab:data_sets}. However, revolution 0999 of \mrk7 has a much lower average count rate (i.e.\ much higher Poisson noise) than the other five observations, introducing considerable scatter to the time lag spectral estimates. Therefore, we recalculate the time lag spectrum for \mrk7 without this observation and the results are shown in the same plot as filled red-diamonds. The time lag spectra in both cases are consistent within the errors, but after the exclusion of revolution 0999 the uncertainties are a good deal smaller. What is very interesting, is that the resultant time lag spectrum of \mrk7 looks very similar to that of \mcg6. In both sources, the time lag spectra are positive at frequencies below $6\times 10^{-4}$ Hz, increasing towards lower frequencies with a power law like shape of slope around unity. At $6\times 10^{-4}$ Hz, the time lags in both sources become negative. They exhibit the same negative minima at around (7$\times 10^{−4}$ Hz, $-18.9\pm4.7$ s) being inconsistent with zero at a 99.99 per cent confidence level. They rise fast again to the positive time lag regime peaking at around ($3\times 10^{-3}$ Hz, 10 s) and at even higher frequencies, they settle to a zero level, although a possible second excursion to the negative time lag regime (at around $10^{-2}$ Hz) is possible.\par
Given these similarities between \mcg6 and \mrk7, after excluding revolution 0999,  we combine the time lag spectra of \mcg6 and \mrk7 to an average time lag spectrum (ATLS hearafter). In the right-hand panel of Fig.~\ref{fig:comb_time_lag} we show the ATLS together with that of \1h0. In the same figure the two solid lines (black and red) correspond to the result of a least square fitting of a quadratic degree B-spline with ten control points to the two datasets, aiming only to outline the basic morphological differences/similarities between the two spectra. In comparison to the ATLS, the time lag spectrum of \1h0 seems to be shifted slightly to higher frequencies, remaining negative at high frequencies and becoming broader around the deepest negative dip. The ATLS is negative over a smaller frequency range, and at higher frequencies exhibits a more pronounced `oscillatory' behaviour.

\subsection{Reverberation and transfer functions}
\label{ssect:reverb_trans_funct}
In this section we attempt to fit the reverberation models that have been previously used for \1h0 \citep{fabian09,miller10_1h0707,zoghbi11} to the ATLS. Since the uncertainties of the ATLS, at high frequencies (greater than $5\times10^{-3}$ Hz), are significantly smaller than those at lower frequencies, any best-fitting model will be driven by these high frequency data. Also above these frequencies the Poisson noise becomes comparable to the intrinsic source variability and the coherence goes significantly below unity (around 0.25) in which case the time lag measurements are no longer meaningful. Therefore, in order to quantify in a statistically robust way the physically interesting part of the observations i.e.\ the part of the time lag spectrum which shows the negative delays and the positive peak at around $3\times 10^{-3}$ Hz, we fit the various models only to the ten lowest frequency points of our data set. Since the initial model-parameter values play a major role for this sort of nonlinear fitting their values are chosen in order to yield an initial model that is phenomenologically in accordance with the data. Note that the following best-fitting model results are only indicative since none of the models possesses a single global minimum in the $\chi^2$ hyper-surface. Also, the quoted errors are simply the diagonal elements of the covariance matrix \citep{bevington92}.\par
In general, the time lag dependence on frequency can be used to constrain reflection models, and in particular the size of the reflector. Following the previous studies of \citet{miller10_1h0707} and \citet{zoghbi11}, we assume that a percentage $f$ of the soft band flux is from a medium which reflects the continuum emission i.e.\ hard band flux, and that the reflector's response to the continuum is uniform over a distribution of time delays, with a width of $\Delta t$, centered at $t_0$ i.e.\ a top hat transfer function (Model 1). Such a model can approximate reflection from a spherical shell, from clumpy material not isotropically surrounding the source, and, to a first approximation, reflection from the disc around the central source. The maximum time delay in this case, $t_{\rm max}=\tau_0+\Delta t/2$, corresponds to the size of the reflector having a radius $ct_{\rm max}/2$. The $\chi^2$ of the best-fitting model is 21.35 for 7 degrees of freedom (d.o.f.) and the best-fitting model parameters are $\Delta t=821\pm236$ s, $t_0=887\pm75$ s and $f=0.13\pm0.02$. This is neither a statistically nor a morphologically acceptable fit, having a hypothesis probability (n.h.p.) of 0.0033 and thus we do no show it.\par
\begin{figure*} 
\includegraphics[width=3in]{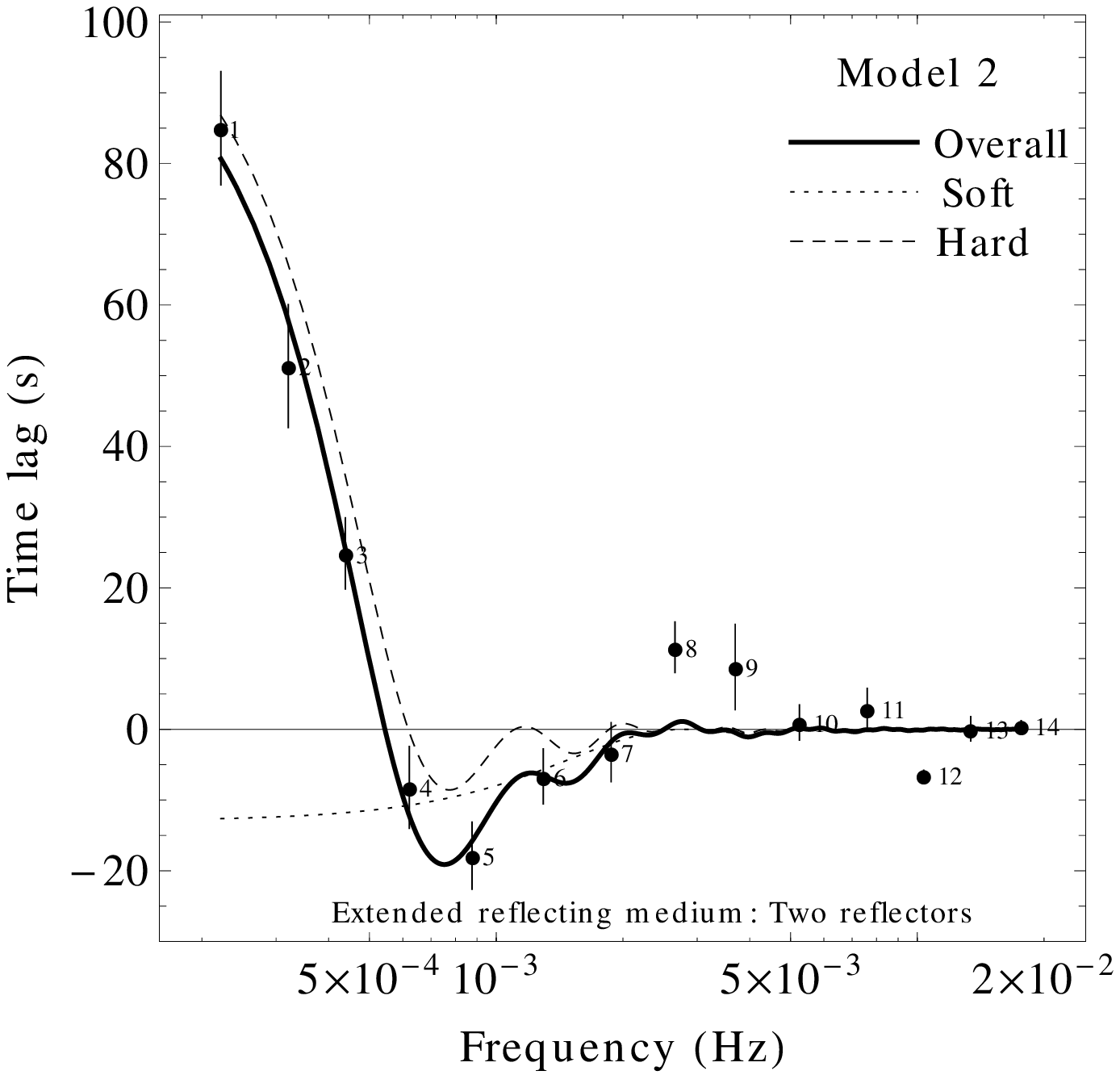}
\includegraphics[width=3in]{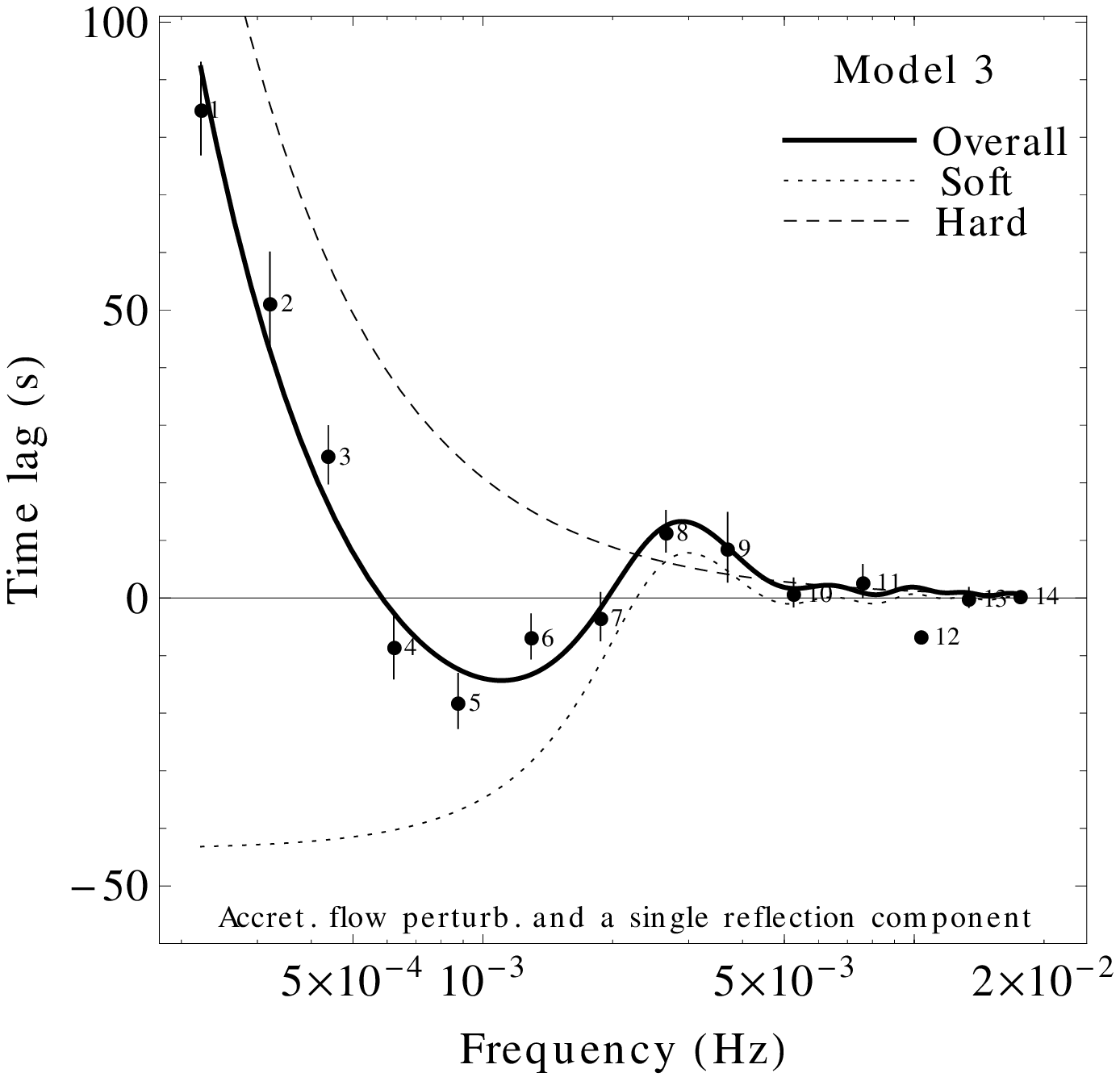}
\caption{Best-fitting models for the ATLS. (Left-hand panel) Model 2 [Double top hat] (black thick solid line) together with the corresponding model components, soft (dotted line) and hard (dashed line). (Right-hand panel) Model 3 [Top hat with power law] (black thick solid line) together with the corresponding model components, soft (dotted line) and hard (dashed line).}
\label{fig:reverb_model} 
\end{figure*}

Then, following \citet{miller10_1h0707}, we considered the possibility that reverberation is likely to be present at differing levels in both the soft and hard bands (Model 2). We therefore considered two reflectors (i.e.\ two top hat transfer functions), of different size (i.e.\ different $\Delta t$), which physically corresponds to an extended reflecting medium, where the hard band photons travel further through the reflecting medium, hence showing longer time delays. In this model the soft band transfer function is assumed to extend from zero to $t_{\rm soft}$ (i.e.\ centered at $t_{\rm soft}/2$) and the hard band transfer function from $t_{\rm soft}$ to $t_{\rm soft}+\Delta t_{\rm hard}$ (i.e.\ centered at $t_{\rm soft}$+$\Delta t_{\rm hard}/2$). The $\chi^2$ of the best-fitting model is 13.18 for 6 d.o.f. (n.h.p=0.040) yielding $t_{\rm soft}=357\pm113$ s, $\Delta t_{\rm hard}=1261\pm105$ s and two relatively small reflection fractions of $f_{\rm soft}= 0.10\pm0.04$ and $f_{\rm hard}= 0.15\pm0.02$. As we can see from the left-hand panel of Fig.~\ref{fig:reverb_model}, Model 2 seems to reproduce adequately the steep transition from the positive to the negative time lag regimes as well as the negative minimum (points 4--7) but it does not reproduce well the subsequent positive peak (points 8--10).\par  
Finally, following \citet{zoghbi11}, we consider a model which the hard band variations are caused by propagation of perturbations in the accretion flow and the soft lag variations are due to a refection component (Model 3). This model consists of a power law model for the positive hard lags and a top hat transfer-function for negative the soft lags. The $\chi^2$ of the best-fitting model is better, 9.63 for 5 d.o.f. (n.h.p=0.0865) yielding for the soft component $t_0=142\pm45$ s, $\Delta t=293\pm47$ s and $f=0.23\pm0.10$ and for the hard component a normalization of $c=(4.4\pm1.8)\times10^{-3}$ s and an index of $q=-1.23\pm0.18$. As we can see from the right-hand panel of Fig.~\ref{fig:reverb_model} Model 3 can very well reproduce the overall shape of the ATLS.

\section{Summary and Discussion}
\label{sect:discus}
Using the longest \textit{XMM-Newton} light curves in the 0.5--1.5 keV and 2--4 keV energy bands we found that \mcg6 and \mrk7 have remarkably similar time lag spectra and above $10^{-3}$ Hz exhibit NXTL similar to those previously observed in \1h0 \citep{fabian09}.\par
With respect to our ATLS, if the NXTL arise through reverberation from scattering of radiation that is passing through partially opaque material (Model 2), the maximum time delay of the order of 1600 s placing a limit on the extent of the reverberating region of approximately 800 light-seconds (assuming a spherical geometry for the scattering medium; this is the most reasonable assumption, even if the covering factor of the material is less than unity). This corresponds for \mcg6 \citep[BH mass $5\times 10^6$ M$_{\sun}$;][]{mchardy05} to 33 $r_{\rm g}$, and for \mrk7 \citep[BH mass $1.8\times 10^6$ M$_{\sun}$;][]{bentz09} to 90 $r_{\rm g}$. Moreover, based on the best-fitting Model 2, the maximum delay for the soft band transfer function is 350 s corresponding to X-rays which are scattered while travelling through a medium, whose opacity decreases with increasing energy. Since, the soft band X-ray photons cannot travel far through the medium before being absorbed any scattered X-rays must have relatively short time delays. This value places an upper limit on the region where the soft X-rays are scattered of 175 s, corresponding to a region of less than 7 $r_{\rm g}$ and 20 $r_{\rm g}$ for \mcg6 and \mrk7 respectively. Thus, even if the NXTL are due to scattering of the X-ray continuum from circumnuclear material, this material must be located within a region very close to the last stable orbit, at least in the case of \mcg6.\par
We note that, the possibility of the continuum being scattered from clouds which are simply located close to the line of sight, but do not cover the sources isotropically, is probably ruled out by the vary fact that we observe NXTL in both \mcg6 and \mrk7. Although the presence of such a specific geometry in one source, i.e.\ \1h0, could be justified as a chance coincidence, its presence in two additional sources renders this possibility rather unlikely.\par
On the other hand, the high frequency shape of the ATLS (negative and then positive) rules out the possibility that there exists just a single time delay between the soft and hard X-ray bands, representing the light travel time between the X-ray source and the reflector. In this case we would expect NXTL at all high frequencies to have the same amplitude, which is not the case in the ATLS (see Fig.~\ref{fig:comb_time_lag}, right-hand panel). The best-fitting Model 3, indicates a source size of the reflector no more than 150 s, corresponding to 6 $r_{\rm g}$ and 17 $r_{\rm g}$ for \mcg6 and \mrk7 respectively.\par
In a top hat reverberation scenario, the wider frequency extent of the NXTL in \1h0 than in the ATLS implies a smaller $\Delta t$ and hence a smaller size of the reflection region. The smaller size implies a smaller BH mass, which is consistent with the overall shift of the lag spectrum to higher frequencies in \1h0. Conversely, the individual time lag spectra of \mrk7 and \mcg6 are very similar; although the admittedly less precisely measured BH mass of \mcg6 is approximately twice that of \mrk7.\par
Regarding possible scalings to XRBs, the zero-crossing frequency in GX\,339-4 is around 4 Hz \citep{uttley11} which is only $5\times 10^3$ times higher than those measured here. The maximum negative lag ($-5\times 10^{-3}$ s for GX\,339-4 vs. -20 s from the ATLS) has a similar scaling, which is again less than the mass scaling, although reverberation may be less important in a hard state XRB.\par
With respect to the power spectral density bends i.e. 7.6$\times10^{-5}$ Hz \citep[\mcg6;][]{mchardy05}, 4.4$\times10^{-4}$ Hz \citep[\mrk7;][]{markowitz07} and 1.6$\times10^{-4}$ Hz \citep[\1h0;][]{zoghbi10} they all seem to occur at lower frequencies where the time lags are strongly positive and they do not seem to scale in any simple manner with any particular frequency in the time lag spectra.\par
Our analysis shows that NXTL are probably a common feature of AGN. However the simple models discussed here provide only an approximate description of the reprocessing mechanism. Therefore it is too early to speculate on the presence of any scaling law in the time lag spectral features.

\section*{Acknowledgments}
DE and IMM acknowledge the Science and Technology Facilities Council (STFC) for support under grant ST/G003084/1. IP acknowledges support by the EU
FP7-REGPOT 206469 grant. Also, we would like to thank Phil Uttley for useful discussions. This research has made use of NASA's Astrophysics Data System Bibliographic Services.

\bibliographystyle{mnauthors}

\bsp
\label{lastpage}
\end{document}